\documentclass[aps,showpacs,nofootinbib,superscriptaddress]{revtex4}
\usepackage{graphicx}
\usepackage{dcolumn}

\begin{document}

\title{ Study of the strong  $\Sigma_c\to \Lambda_c\, \pi$, 
$\Sigma_c^{*}\to \Lambda_c\, \pi$ and
$\Xi_c^{*}\to \Xi_c\, \pi$ decays in a nonrelativistic quark model}
\author{C. Albertus} \affiliation{Departamento de
F\'{\i}sica Moderna, Universidad de Granada, E-18071 Granada, Spain.}
\author{ E. Hern\'andez} \affiliation{Grupo de F\'\i sica Nuclear,
Facultad de Ciencias, E-37008 Salamanca, Spain.}  
\author {J. Nieves}
\affiliation{Departamento de F\'{\i}sica Moderna, Universidad de
Granada, E-18071 Granada, Spain.}
\author{ J. M. Verde-Velasco} \affiliation{Grupo de F\'\i sica Nuclear,
Facultad de Ciencias, E-37008 Salamanca, Spain.} 
\begin{abstract} 
\rule{0ex}{3ex} 

\end{abstract}


\begin{abstract} 
 We present  results for   the strong widths corresponding to  the
 $\Sigma_c\to \Lambda_c\, \pi$,
$\Sigma_c^{*}\to \Lambda_c\, \pi$ and
$\Xi_c^{*}\to \Xi_c\, \pi$ decays. The calculations
have been done  in a nonrelativistic constituent quark model with wave
functions that take advantage of the constraints imposed by heavy quark
symmetry. Partial conservation of axial current hypothesis allows us to  
 determine the  strong vertices from an analysis of the axial current matrix
elements. Our results  \hbox{$\Gamma(\Sigma_c^{++}\to \Lambda_c^+\, \pi^+)=2.41
\pm0.07\pm0.02\,\mathrm{MeV}$},
{$\Gamma(\Sigma_c^{+}\to \Lambda_c^+\, \pi^0)=2.79
\pm0.08\pm0.02\,\mathrm{MeV}$},
{$\Gamma(\Sigma_c^{0}\to \Lambda_c^+\, \pi^-)=2.37
\pm0.07\pm0.02\,\mathrm{MeV}$},
{$\Gamma(\Sigma_c^{*\,++}\to \Lambda_c^+\, \pi^+)=17.52\pm0.74\pm0.12\,
\mathrm{MeV}$}, 
{$\Gamma(\Sigma_c^{*\,+}\to \Lambda_c^+\, \pi^0)=17.31\pm0.73\pm0.12\,
\mathrm{MeV}$}, 
{$\Gamma(\Sigma_c^{*\,0}\to \Lambda_c^+\, \pi^-)=16.90\pm0.71\pm0.12\,
\mathrm{MeV}$}, 
{$\Gamma(\Xi_c^{*\,+}\to \Xi_c^0\, \pi^++\Xi_c^+\pi^0)=3.18\pm0.10\pm0.01
\,\mathrm{MeV}$} and
{$\Gamma(\Xi_c^{*\,0}\to \Xi_c^+\, \pi^-+\Xi_c^0\pi^0)=3.03\pm0.10\pm0.01
\,\mathrm{MeV}$}
are in
good agreement with experimental determinations.
\end{abstract}
\maketitle

\section{Introduction}
The nonrelativistic constituent quark model (NRCQM), using QCD-inspired
potentials, has proved to be an excellent tool to predict properties of hadrons.
In the case of baryons including one heavy quark $c$ or $b$ and two light ones
$u$, $d$ or $s$, we can take advantage of yet another property of QCD: Heavy
quark symmetry (HQS)~\cite{nussinov,voloshin,politzer,iw}. This symmetry  
arises when the heavy quark  mass is much larger than the QCD scale
($\Lambda_{QCD}$).
In that limit the dynamics of the light quark degrees of freedom becomes
independent of the heavy quark flavor and spin.
The light degrees of freedom are thus well defined and the masses of the baryons depend only on the quark content and on the
light-light quantum numbers.
This simplification was used in Ref.~\cite{albertus04}
to solve the three-body problem for the ground state (L=0) of baryons with a heavy quark
using a simple variational ansazt. We obtained static properties 
(masses, mass and charge radii\dots) in good
agreement with previous calculations that used more involved Faddeev
equations~\cite{silvestre96}. The advantage of our approach is that we also
obtain easy to handle wave functions. Those wave functions were already used
in Ref.~\cite{albertus05} to study, with  good results, the  semileptonic decays of $\Lambda_b$
and $\Xi_b$ baryons.

In this work we shall evaluate strong widths for the $\Sigma_c\to
\Lambda_c\, \pi$, $\Sigma_c^{*}\to \Lambda_c\, \pi$ and $\Xi_c^{*}\to
\Xi_c\, \pi$ decays.  Last decade has ~seen a great progress on
charmed baryon physics and now the ground state baryons with a $c$
quark, with the exception of the $\Omega_c^*$, are well
established~\cite{pdg04}, and we have experimental information on the
strong one-pion decay widths for the
$\Sigma_c$~\cite{cleo01,cleo02,focus02},
$\Sigma_c^*$~\cite{cleo01,cleo05} and $\Xi_c^*$~\cite{cleo96,cleo95}.
With very little kinetic energy available in the final state these
reactions should be well described in a nonrelativistic approach. 
Although they have been analyzed before in the framework of the
constituent quark model (CQM)~\cite{rosner95,pirjol97},  no attempt was made
there to evaluate the full matrix elements. While there have been dynamical
calculations in other models (see references below), to our knowledge, ours is the first 
dynamical calculation within a nonrelativistic approach.
In our calculation we will use the HQS--constrained wave functions that we
evaluated in Ref.~\cite{albertus04} using different inter-quark interactions,
 and  whose goodness have already been tested in
the study of the semileptonic $\Lambda_b\to \Lambda_c$ and $\Xi_b\to\Xi_c$ 
decays in  Ref.~\cite{albertus05}. The use of different quark--quark potentials
 will allow us
 to obtain theoretical  uncertainties on the widths due to the 
 quark-quark interaction. The pion emission amplitude will
be obtained in a spectator model (one-quark pion emission) with the
use of partial conservation of axial current hypothesis (PCAC).

These reactions, and similar ones, have also been  addressed  in 
QCD sum rules (QCDSR)~\cite{grozin98,zhu98}, in
heavy hadron chiral 
perturbation theory (HHCPT)~\cite{pirjol97,yan92,huang95,cheng97,chiladze97}, 
and  within 
relativistic quark models like the light-front quark
model (LFQM)~\cite{tawfiq98} and the  relativistic three-quark
model (RTQM)~\cite{ivanov9899}.

\section{One-pion emission amplitude ${\cal A}_{BB'\pi}^{(s,s')}(P_{B},P_{B'})$}
\label{sect:sdwpcac}
To determine the pion emission amplitude ${\cal A}_{BB'\pi}^{(s,s')}(P_{B},P_{B'})$
we shall
use PCAC, as we have done 
 in the meson sector, in a previous study of the strong
$B^*B\pi$ and $D^*D\pi$ couplings~\cite{albertus05-2}.
PCAC allows us to relate that amplitude to the matrix element of 
the divergence of the axial current.
For the emission of a $\pi^+$ we have\footnote{Note that we give the
expression corresponding to the non--pole part of the matrix element. If
 the pion pole contribution is included then the relation is given by
$
\left\langle B',\ s'\,\vec{P}_{B'}\,\left|\,q^\mu\, J_{A\,\mu}^{d\,u}(0)
\,\right|
\,B,\ s\,\vec{P}_B\,\right\rangle= -i\,f_{\pi}\frac{m_{\pi}^2}{q^2-m_{\pi}^2}\ 
{\cal A}^{(s,s')}_{BB'\pi^+}(P_B,P_{B'})
$.}
\begin{eqnarray}
\label{np}
\left\langle B',\ s'\,\vec{P}_{B'}\,|\,q^\mu\, J_{A\,\mu}^{d\, u}(0)\,|
\,B,\ s\,\vec{P}_B\,\right\rangle_{non-pole}&=& i\,f_{\pi}\ 
{\cal A}^{(s,s')}_{BB'\pi^+}(P_B,P_{B'})\end{eqnarray}
where $s,\, s'$ are 
the third component of the
spin of the   $B,\, B'$ baryons in their respective center of mass 
systems,  $P_B=(E_B(|\vec{P}_B|),\,\vec{P}_B),\, P_{B'}=(
E_{B'}(|\vec{P}_{B'}|),\,\vec{P}_{B'}) $ are their respective 
four--momenta  and $q=P_B-P_{B'}$. $J_{A\,\mu}^{d\, u}(0)$ is the axial current for the $u\to d$ transition,
and $f_{\pi}=130.7
$\,MeV~\cite{pdg04} is the pion decay constant. The baryon states are
normalized as $\left\langle B, s' \vec{P}'\,| B, s \vec{P}
\right\rangle=\delta_{s,s'}\,(2\pi)^3\,2E_B(|\vec{P}\,|)\,
\delta^3(\vec{P}-\vec{P}')$. Furthermore we shall use physical masses
taken from Ref.~\cite{pdg04} in all calculations.
\section{Description of baryon states}
\label{sect:bs}
We use the following expression for the state of a baryon $B$ 
with three-momentum $\vec{P}$ and spin projection $s$ 
 in the baryon center of mass
\begin{eqnarray}
\label{wf}
&&\hspace{-1cm}\left|{B,s\,\vec{P}}\,\right\rangle_{NR}
=\int d^{\,3}Q_1 \int d^{\,3}Q_2\ \frac{1}{\sqrt2}\sum_{\alpha_1,\alpha_2,\alpha_3}
\hat{\psi}^{(B,s)}_{\alpha_1,\alpha_2,\alpha_3}(\,\vec{Q}_1,\vec{Q}_2\,)
\ \frac{1}{(2\pi)^3\ \sqrt{2E_{f_1}(|\vec{p}_1|)2E_{f_2}(|\vec{p}_2|)
2E_{f_3}(|\vec{p}_3|)}}\nonumber\\ 
&&\hspace{3cm}
\times\left|\ \alpha_1\
\vec{p}_1=\frac{m_{f_1}}{\overline{M}}\vec{P}+\vec{Q}_1\ \right\rangle
\left|\ \alpha_2\ \vec{p}_2=\frac{m_{f_2}}{\overline{M}}\vec{P}+\vec{Q}_2\ \right\rangle
\left|\ \alpha_3\ \vec{p}_3=\frac{m_{f_3}}{\overline{M}}\vec{P}-\vec{Q}_1
-\vec{Q}_2\ \right\rangle
 \end{eqnarray}
$\alpha_j$ 
represents
the quantum numbers of spin $s$, flavor $f$ and color $c$ 
($\alpha_j\equiv(s_j,f_j,c_j)$) of the {\it j-th} quark,
 while $(E_{f_j}(|\vec{p}_j|),\,\vec{p}_j)$ and $m_{f_j}$ represent its four--momentum 
 and   mass. $\overline{M}$ stands for $\overline{M}=m_{f_1}+m_{f_2}+m_{f_3}$. We choose the
 third quark to be the $c$ quark while the first two  will be the light
 ones. 
 The normalization of the quark 
states is 
$
\left\langle\ \alpha^{\prime}\ \vec{p}^{\ \prime}\,|\,\alpha\
\vec{p}\, \right\rangle=\delta_{\alpha^{\prime},\ \alpha}\, (2\pi)^3\,
2E(|\vec{p}\,|)\,\delta^3( \vec{p}^{\ \prime}-\vec{p}\,)
$. 
Besides, $\hat{\psi}^{\,(B,s)}_{\alpha_1,\alpha_2,\alpha_3}
(\,\vec{Q}_1,\vec{Q}_2\,)$ is the nonrelativistic
momentum space wave
function  for the internal motion, being
$\vec{Q}_1$ and $\vec{Q}_2$  the momenta conjugate  to the relative positions $\vec{r}_1$ and 
$\vec{r}_2$ of the
two light quarks  with respect to the heavy one. This
wave function is
  antisymmetric under the simultaneous exchange $\alpha_1\longleftrightarrow
\alpha_2, \vec{Q}_1 \longleftrightarrow \vec{Q}_2 $, being 
    also antisymmetric under an overall exchange of
the color degrees of freedom. It is normalized such that 
\begin{equation}
\int d^{\,3}Q_1 \int d^{\,3}Q_2\ \sum_{\alpha_1,\alpha_2,\alpha_3}
\left(\hat{\psi}^{(B,s')}_{\alpha_1,\alpha_2,\alpha_3}(\,\vec{Q}_1,\vec{Q}_2\,)\right)^*
\hat{\psi}^{(B,s)}_{\alpha_1,\alpha_2,\alpha_3}(\,\vec{Q}_1,\vec{Q}_2\,)
=\delta_{s',\, s}
\end{equation}
and, thus,  the normalization of our nonrelativistic baryon states is 
\begin{equation}
{}_{\stackrel{}{NR}}\left\langle\, {B,s'\,\vec{P}^{\,\prime}}\,|\,{B,s
\,\vec{P}}\,\right\rangle_{NR}
=\delta_{s',\,s}\,(2\pi)^3\,\delta^3(\vec{P}^{\,\prime}-\vec{P}\,)
\end{equation}
For the particular case of ground state  $\Lambda_c$, $\Sigma_c$,
$\Sigma^{*}_c$, $\Xi_c$ and $\Xi^{*}_c$ we can assume the
orbital angular momentum to be zero. We will also take advantage of HQS and
assume the light--degrees of freedom quantum numbers are well defined (For
quantum numbers see, 
for instance, Table 1 in Ref~\cite{albertus04}). In that case we have\footnote{We only give the wave
function for the baryons involved in $\pi^+$ emission. Wave functions for
other isospin states 
of the same baryons are
easily constructed.}
\begin{eqnarray}
\hat{\psi}^{\,(\Lambda^+_c, s)}_{\alpha_1,\,\alpha_2,\,\alpha_3}(\,\vec{Q}_1,\vec{Q}_2\,)
&=&\frac{\varepsilon_{c_1\,c_2\,c_3}}{\sqrt{3}!}
\ (1/2,1/2,0;s_1,s_2,0)\nonumber\\
&&\hspace{2cm}\times
\frac{\delta_{f_3,\,c}\,\delta_{s_3,\,s}}{\sqrt2}
\left(
\delta_{f_1,\,u}\,\delta_{f_2,\,d}\,\widetilde{\phi}^{S_{l}=0}_{u,\,d,\,c}(\,\vec{Q}_1,\vec{Q}_2\,)\ 
-\delta_{f_1,\,d}\,\delta_{f_2,\,u}\,\widetilde{\phi}^{S_{l}=0}_{d,\,u,\,c}(\,\vec{Q}_1,\vec{Q}_2\,)\
\right)  
 \nonumber\\
\hat{\psi}^{\,(\Sigma^{++}_c,s)}_{\alpha_1,\,\alpha_2,\,\alpha_3}(\,\vec{Q}_1,\vec{Q}_2\,)
&=&\frac{\varepsilon_{c_1\,c_2\,c_3}}{\sqrt{3}!}\
\ \widetilde{\phi}^{S_{l}=1}_{u,\,u,\,c}(\,\vec{Q}_1,\vec{Q}_2\,)\ 
\delta_{f_1,\,u}\, \delta_{f_2,\,u}\, \delta_{f_3,\,c}
\sum_m (1/2,1/2,1;s_1,s_2,m)\ 
(1,1/2,1/2;m,s_3,s)\nonumber\\
\hat{\psi}^{\,(\Sigma^{*\,++}_c,s)}_{\alpha_1,\,\alpha_2,\,\alpha_3}(\,\vec{Q}_1,\vec{Q}_2\,)
&=&\frac{\varepsilon_{c_1\,c_2\,c_3}}{\sqrt{3}!}\
\ \widetilde{\phi}^{S_{l}=1}_{u,\,u,\,c}(\,\vec{Q}_1,\vec{Q}_2\,)\ 
\delta_{f_1,\,u}\,\delta_{f_2,\,u}\, \delta_{f_3,\,c}
\ 
\sum_m (1/2,1/2,1;s_1,s_2,m)\ 
(1,1/2,3/2;m,s_3,s)\nonumber\\
\hat{\psi}^{\,(\Xi^0_c,s)}_{\alpha_1,\,\alpha_2,\,\alpha_3}(\,\vec{Q}_1,\vec{Q}_2\,)
&=&\frac{\varepsilon_{c_1\,c_2\,c_3}}{\sqrt{3}!}
\ 
(1/2,1/2,0;s_1,s_2,0)\nonumber\\
&&\hspace{2cm}\times 
\frac{\delta_{s_3,\,s}\,\delta_{f_3,\,c} }{\sqrt2}\, \left(
\delta_{f_1,\,d}\,\delta_{f_2,\,s}\ \widetilde{\phi}^{S_{l}=0}_{d,s,c}(\,\vec{Q}_1, \vec{Q}_2\,)
-\delta_{f_1,\,s}\,\delta_{f_2,\,d}\ \widetilde{\phi}^{S_{l}=0}_{s,d,c}(\,\vec{Q}_1, \vec{Q}_2\,)
\right)\nonumber\\
\hat{\psi}^{\,(\Xi^{*\,+}_c,s)}_{\alpha_1,\,\alpha_2,\,\alpha_3}(\,\vec{Q}_1,\vec{Q}_2\,)
&=&\frac{\varepsilon_{c_1\,c_2\,c_3}}{\sqrt{3}!}
\ \sum_m (1/2,1/2,1;s_1,s_2,m)\,(1,1/2,3/2;m,s_3,s)
\nonumber\\
&&\hspace{2cm} \times\ \frac{\delta_{f_3,\,c}}{\sqrt2}\,\left(
\delta_{f_1,\,u}\,\delta_{f_2,\,s}\ \widetilde{\phi}^{S_{l}=1}_{u,s,c}(\,\vec{Q}_1, \vec{Q}_2\,)
+\delta_{f_1,\,s}\,\delta_{f_2,\,u}\ \widetilde{\phi}^{S_{l}=1}_{s,u,c}(\,\vec{Q}_1, \vec{Q}_2\,)
\right)
\end{eqnarray}
Here $\varepsilon_{c_1 c_2 c_3}$ is the fully antisymmetric tensor on color
indices being $\varepsilon_{c_1 c_2 c_3}/\sqrt{3!}$  the antisymmetric color wave
function, the $(j_1,j_2,j;m_1,m_2,m_3)$  are Clebsch-Gordan coefficients and 
the
$\widetilde{\phi}^{S_l}_{f_1,\,f_2,\,f_3}(\,\vec{Q}_1, \vec{Q}_2\,)$, with
$S_l$ the total spin of the light degrees
of freedom, are the  Fourier transform of the corresponding normalized coordinate space 
wave functions obtained in  Ref.~\cite{albertus04}.
Their dependence on momenta is through  $|\vec{Q}_1|$, $|\vec{Q}_2|$ 
and $\vec{Q}_1\cdot\vec{Q}_2$ alone, and they are symmetric under the
simultaneous exchange $f_1\longleftrightarrow
f_2, \vec{Q}_1 \longleftrightarrow \vec{Q}_2 $. 
\section{Results}
\subsection{$\Sigma_c\to\Lambda_c \,\pi$ decay}
\label{sect:sig}
Let us do the $\Sigma_c^{++}\to\Lambda_c^+ \,\pi^+$ case. 
The matrix element of the divergence $q^\mu J_{A\,\mu}^{d\, u}(0)$ of  the axial 
 current determines the $\pi^+$ emission  amplitude as
\begin{eqnarray}
\label{eq:amsiglam}
{\cal A}^{(s,s')}_{\Sigma_c^{++}\Lambda_c^+\pi^+}(P,P')=\frac{-i}{f_\pi}\ 
\left\langle \Lambda_c^+,\ s'\,\vec{P}'\,\left|\,q^\mu\, J_{A\,\mu}^{d\, u}(0)
\,\right|
\,\Sigma_c^{++},\ s\,\vec{P}\,\right\rangle_{non-pole}
=ig_{\Sigma_c^{++}\Lambda_c^+\pi^+}
\ \overline{u}_{\Lambda_c^+\ s'}(\vec{P}'\,)\
\gamma_5\
u_{\Sigma_c^{++}\ s}(\vec{P})
\end{eqnarray}
where the coupling constant $g_{\Sigma_c^{++}\Lambda_c^+\pi^+}$,
 in analogy to the pion coupling to nucleons and nucleon resonances,
has been chosen to be dimensionless, and $u_{\Sigma_c^{++}\ s}(\vec{P})$,
  $u_{\Lambda_c^{+}\ s'}(\vec{P}')$ are Dirac spinors
 normalized to twice the energy.  
 The width is given by
\begin{equation}
\label{eq:gsiglam}
\Gamma(\Sigma_c^{++}\to\Lambda_c^+\pi^+)=\frac{|\vec{q}\,|}{8\pi M_{\Sigma_c^{++}}^2 }\
g_{\Sigma_c^{++}\Lambda_c^+\pi^+}^2\left(\ (
M_{\Sigma_c^{++}}-M_{\Lambda_c^+})^2-m^2_{\pi}\right)
\end{equation}
with $|\vec{q}\,|$ the modulus of the final baryon or pion three-momentum. From Eq.~(\ref{eq:amsiglam}), taking
$\vec{P}=\vec{0}$,  $\vec{P}'=-|\vec{q}\,|\,\vec{k}$ in the $z$ direction,
$s=s'=1/2$, and taking into account the different normalization of our
nonrelativistic states, we have 
\begin{eqnarray}
\hspace{-.5cm}g_{\Sigma_c^{++}\Lambda_c^+\pi^+}
&=&\frac{-1}{f_\pi}\ 
\frac{\sqrt{E_{\Lambda_c^+}(|\vec{q}\,|)+M_{\Lambda_c^+}}
\sqrt{2M_{\Sigma_c^{++}} 2E_{\Lambda_c^+}(|\vec{q}\,|) }}{|\vec{q}\,|
\sqrt{2M_{\Sigma_c^{++}} }}
\left(
(M_{\Sigma_c^{++}}- E_{\Lambda_c^+}(|\vec{q}\,|) )\ A_{\Sigma_c^{++}\Lambda_c^+
,\ 0}^{1/2,1/2}
+|\vec{q}\,|\ A_{\Sigma_c^{++}\Lambda_c^+
,\ 3}^{1/2,1/2}
\right)
\end{eqnarray}
with
\begin{eqnarray}
A_{\Sigma_c^{++}\Lambda_c^+
,\ \mu}^{1/2,1/2}={}_{\stackrel{}{NR}}\left\langle \Lambda_c^+,\
1/2\ -|\vec{q}|\,\vec{k}\ \left|\  J_{A\,\mu}^{d\, u}(0)\ \right|
\,\Sigma_c^{++},\ 1/2\ \vec{0}\,\right\rangle_{NR,\ non-pole}
\end{eqnarray}
The $A_{\Sigma_c^{++}\Lambda_c^+
,\ \mu}^{1/2,1/2}$  are easily evaluated using one-body current operators and their 
expressions  can be found in the appendix.
\begin{table}[t]
\begin{center}
\begin{tabular}{l|c c c c}
\hline
 & \hspace{.25cm}$g_{\Sigma_c^{++}\Lambda_c^+\pi^+}$\hspace{.25cm} &
 \hspace{.25cm}$\Gamma(\Sigma_c^{++}\to\Lambda_c^+\pi^+)$ \hspace{.25cm} &
 \hspace{.25cm} $\Gamma(\Sigma_c^{+}\to\Lambda_c^+\pi^0)$ \hspace{.25cm} &
 \hspace{.25cm}$\Gamma(\Sigma_c^{0}\to\Lambda_c^+\pi^-)$ \hspace{.25cm} 
 \\
 &&[MeV]&[MeV]&[MeV]\\
\hline
This work 
& $21.73\pm0.32\pm0.08$&$2.41\pm0.07\pm0.02$ &$2.79\pm0.08\pm0.02$ 
&$2.37\pm0.07\pm0.02$\\
\hline
Experiment &  & $2.3\pm0.2\pm 0.3$~\cite{cleo02} & $< 4.6$ (CL=90\%)~\cite{cleo01} & 
$2.5\pm0.2\pm0.3$~\cite{cleo02} \\
 &  &$2.05^{+0.41}_{-0.38}\pm0.38$~\cite{focus02} & &
 $1.55^{+0.41}_{-0.37}\pm0.38$~\cite{focus02}\\
 \hline
Theory & &  & &\\
CQM
& & $1.31\pm 0.04$~\cite{rosner95} &$1.31\pm 0.04$~\cite{rosner95}& 
$1.31\pm 0.04$~\cite{rosner95}\\
 & &$2.025^{+1.134}_{-0.987}$~\cite{pirjol97} & &
 $1.939^{+1.114}_{-0.954}$~\cite{pirjol97} \\
HHCPT &22, 29.3~\cite{yan92} & 2.47, 4.38~\cite{yan92}& 2.85,
5.06~\cite{yan92}&2.45, 4.35~\cite{yan92}\\
 & & 2.5~\cite{huang95}&  3.2~\cite{huang95}& 2.4~\cite{huang95}\\
 & & &  &$1.94\pm0.57$~\cite{cheng97}\\
LFQM & & 1.64 ~\cite{tawfiq98}&1.70 ~\cite{tawfiq98} & 1.57 ~\cite{tawfiq98}\\
RTQM & &$2.85\pm0.19$~\cite{ivanov9899} &$3.63\pm0.27$~\cite{ivanov9899}
  & $2.65\pm0.19$~\cite{ivanov9899} \\
\hline
\end{tabular}
\end{center}
\caption{ Coupling constant  $g_{\Sigma_c^{++}\Lambda_c^+\pi^+}$ and 
total widths $\Gamma(\Sigma_c^{++}\to\Lambda_c^+\pi^+)$,
$\Gamma(\Sigma_c^{+}\to\Lambda_c^+\pi^0)$ and
$\Gamma(\Sigma_c^{0}\to\Lambda_c^+\pi^-)$
(See text for details).
  Experimental data and different theoretical
calculations are also shown.}
\label{tab:siglam}
\end{table}

In Table~\ref{tab:siglam} we present the results for 
$g_{\Sigma_c^{++}\Lambda_c^+\pi^+}$ and the widths 
$\Gamma(\Sigma_c^{++}\to\Lambda_c^+\pi^+)$,
$\Gamma(\Sigma_c^{+}\to\Lambda_c^+\pi^0)$ and 
$\Gamma(\Sigma_c^{0}\to\Lambda_c^+\pi^-)$. 
To get the  values for 
$\Gamma(\Sigma_c^{+}\to\Lambda_c^+\pi^0)$ and
$\Gamma(\Sigma_c^{0}\to\Lambda_c^+\pi^-)$ we use 
$g_{\Sigma_c^{++}\Lambda_c^+\pi^+}$ and make the appropriate mass changes in the rest
of factors in Eq.~(\ref{eq:gsiglam}).
 Our results show two types of errors. The second one
 results from the Monte Carlo evaluation of the integrals needed to obtain the
 $g_{\Sigma_c^{++}\Lambda_c^+\pi^+}$ coupling constant. The first one, can be
 considered as a theoretical uncertainty and comes from the use of different quark--quark
interactions \footnote{
We use five different inter-quark interactions, 
 one suggested by Bhaduri and
collaborators~\cite{bhaduri81}, and four others suggested 
by Silvestre-Brac and Semay
~\cite{silvestre96,sbs93}. All of them contain a 
confinement term, plus
  Coulomb  and  hyperfine terms coming from one-gluon exchange, and differ
  from one another in the form factors used for the
hyperfine terms, the power of the confining term  or
the use of a form factor in the one gluon exchange Coulomb potential.}.
The results are in very good agreement with the experimental data by the CLEO
Collaboration in Refs.~\cite{cleo02,cleo01}. The value for
 $\Gamma(\Sigma_c^{++}\to\Lambda_c^+\pi^+)$ also agrees with the experimental
 data by the FOCUS Collaboration in Ref.~\cite{focus02}. The agreement with FOCUS
 data is not  good for the $\Gamma(\Sigma_c^{0}\to\Lambda_c^+\pi^-)$ case,
 although our result is still within experimental errors. 
 Our results show variations as large as $\approx$17\%      between different charge configurations. This is due to the little
 kinetic energy available in the final state that makes the widths very
 sensitive to the precise masses of the hadrons involved.
In this respect there is a new precise determination of the
 $\Lambda_c^+$ mass by the {\sl BABAR} collaboration
 $M_{\Lambda_c}=2286.46\pm0.14$\ MeV/c$^2$~\cite{babar05}, which is roughly 1.5
 MeV/c$^2$ above the value quoted by the Particle Data Group in 
 Ref.~\cite{pdg04}. With this new value our calculated widths would get 
 reduced by $9\%
 $. This reduction comes from  phase space factors while the coupling 
 $g_{\Sigma_c^{++}\Lambda_c^+\pi^+}$ changes only at the level of
 0.1\%.\\
 As for the
 other theoretical determinations, the CQM calculation in Ref.~\cite{rosner95} uses exact $s \longleftrightarrow c$
symmetry to relate the $\Sigma_c\to \Lambda_c\pi$ decay to the non--charmed 
 $\Sigma^*\to \Lambda\pi$ analogue decay.  Their results are smaller than the
 experimental data obtained by the CLEO Collaboration. In
 Ref.~\cite{pirjol97} a unique coupling constant is fixed in order to reproduce
 all experimental information on $\Sigma_c^*\to\Lambda_c\pi$ widths. That
 coupling is latter used to predict the
 $\Sigma_c\to\Lambda_c\pi$ widths. This coupling suffers from large
 uncertainties and thus the theoretical errors on the predicted widths
 are also very large. In the HHCPT calculation of Ref.~\cite{yan92}  a
 simple CQM argument is used in order to obtain the unknown coupling constant in
 the HHCPT Lagrangian. Furthermore the 
 authors allow for a renormalization of the axial coupling $g_A^{ud}$ for light 
 quarks. The
 largest of the two values quoted corresponds to the case where that coupling
 is unrenormalized and then is given by $g_A^{ud}=1$. The smaller number quoted
 corresponds to the use of a renormalized value of $g_A^{ud}=0.75$. The case $g_A^{ud}=1$ is the one that compares with our
 calculation. Their results for the widths almost double ours and are not in
 agreement with experiment. Their simple determination of the coupling constant
 can not be correct. The values obtained in the HHCPT calculation of
 Ref.~\cite{huang95} are closer to our results and experimental data. There the 
 authors determine
 the needed coupling constants from the analysis of analogue decays involving
 non--charmed baryons. In Ref.~\cite{cheng97}, also within the HHCPT approach,
 and similarly to Ref.~\cite{pirjol97},
 the authors fix the unknown coupling in the Lagrangian using the experimental
 data for the $\Sigma_c^*\to\Lambda_c\pi$ decays. From there they predict the
  $\Gamma(\Sigma_c^0\to\Lambda_c^+\pi^-)$ obtaining a value very close to the
 one in Ref.~\cite{pirjol97}, and that  suffers also  from large uncertainties.
 The two relativistic quark model calculations of
 Refs.~\cite{tawfiq98,ivanov9899} give results that differ by almost a factor
 of two. Our results are closer to the ones obtained within the RTQM
of Ref.~\cite{ivanov9899}.
\subsection{$\Sigma_c^{*}\to \Lambda_c\pi$  decay}
\label{sect:sig*}
Let us  analyze the case with a $\pi^+$ in the final state,
$\Sigma_c^{*\,++}\to \Lambda_c^+\pi^+$.
Similarly to the $\Sigma_c$ decay  before we now have
\begin{eqnarray}
\label{eq:amsig*lam}
{\cal A}^{(s,s')}_{\Sigma_c^{*\,++}\Lambda_c^+\pi^+}(P,P')&=&\frac{-i}{f_\pi}
\left\langle \Lambda_c^+,\ s'\,\vec{P}'\,\left|\,q^\mu\, J_{A\,\mu}^{d\, u}(0)
\,\right|
\,\Sigma_c^{*\,++},\ s\,\vec{P}\,\right\rangle_{non-pole}\nonumber\\
&=&i\frac{g_{\Sigma_c^{*\,++}\Lambda_c^+\pi^+}}{2M_{\Lambda_c^+}}
\ q_\nu\ \overline{u}_{\Lambda_c^+\ s'}(\vec{P}'\,)\
u_{\Sigma_c^{*\,++}\ s}^\nu(\vec{P})
\end{eqnarray}
where we have introduced the dimensionless coupling constant
$g_{\Sigma_c^{*\,++}\Lambda_c^+\pi^+}$
and $u_{\Sigma_c^{*\,++}\ s}^\nu(\vec{P})$ is a Rarita-Schwinger
spinor normalized to twice the energy.
The width is given by
\begin{equation}
\label{eq:gammasig*lam}
\Gamma(\Sigma_c^{*\,++}\to\Lambda_c^+\pi^+)=\frac{|\vec{q}\,|^3}{24\pi M_{\Sigma_c^{*\,++}}^2 }\
\frac{g_{\Sigma_c^{*\,++}\Lambda_c^+\pi^+}^2}{4M_{\Lambda_c^{+}}^2} \left((
M_{\Sigma_c^{*\,++}}+M_{\Lambda_c^+})^2-m^2_{\pi}\right)
\end{equation}
\begin{table}[t]
\begin{center}
\begin{tabular}{l|c c c c}
\hline
 & \hspace{.25cm}$g_{\Sigma_c^{*\, ++}\Lambda_c^+\pi^+}$\hspace{.25cm}
 & \hspace{.25cm}$\Gamma(\Sigma_c^{*\,++}\to\Lambda_c^+\pi^+)$ \hspace{.25cm}
 & \hspace{.25cm}$\Gamma(\Sigma_c^{*\,+}\to\Lambda_c^+\pi^0)$ \hspace{.25cm}
 & \hspace{.25cm}$\Gamma(\Sigma_c^{*\,0}\to\Lambda_c^+\pi^-)$ \hspace{.25cm}\\
 &&[MeV]&[MeV]&[MeV]\\
\hline
This work 
&$36.20\pm0.75\pm0.13$&$17.52\pm0.74\pm0.12$&$17.31\pm0.73\pm0.12$ &
$16.90\pm0.71\pm0.12$\\
\hline
Experiment &  & $14.1^{+1.6}_{-1.5}\pm 1.4$~\cite{cleo05}& $< 17$ (CL=90\%)~\cite{cleo01} & $16.6^{+1.9}_{-1.7}\pm1.4$~\cite{cleo05}\\
\hline
Theory &&&&\\
QCDSR &$13.8\div 24.2$~\cite{grozin98}&&&\\
&$32.5\pm2.1\pm6.9$~\cite{zhu98} & & &\\
CQM& & 20~\cite{rosner95}&20~\cite{rosner95} &20~\cite{rosner95}\\
HHCPT   & & 25~\cite{huang95}&25~\cite{huang95} &25~\cite{huang95}\\
LFQM   & & 12.84~\cite{tawfiq98} & & 12.40~\cite{tawfiq98}\\
RTQM& &$21.99\pm0.87$~\cite{ivanov9899} &
  & $21.21\pm0.81$~\cite{ivanov9899} \\
\hline
\end{tabular}
\end{center}
\caption{ Coupling constant $g_{\Sigma_c^{*\,++}\Lambda_c^+\pi^+}$ and 
total widths $\Gamma(\Sigma_c^{*\,++}\to\Lambda_c^+\pi^+)$,
$\Gamma(\Sigma_c^{*\,+}\to\Lambda_c^+\pi^0)$ and
$\Gamma(\Sigma_c^{*\,0}\to\Lambda_c^+\pi^-)$. 
 Experimental data and different theoretical
calculations are also shown. Note that in order to compare with our definition
of $g_{\Sigma_c^{*\,++}\Lambda_c^+\pi^+}$ we have multiplied the coupling
constants evaluated in Refs.~\cite{grozin98,zhu98} by $2M_{\Lambda_c^+}/f_\pi$.
}
\label{tab:sig*lam}
\end{table}
Taking again 
$\vec{P}=\vec{0}$, $\vec{P}'=-|\vec{q}\,|\,\vec{k}$ in the $z$ direction,
 and $s=s'=1/2$ we obtain from Eq.(\ref{eq:amsig*lam})
 \begin{eqnarray}
\label{eq:gsig*lam}
\hspace{-0.5cm}g_{\Sigma_c^{*\,++}\Lambda_c^+\pi^+}
&=&\frac{\sqrt3}{f_\pi\sqrt2}\ \frac{2M_{\Lambda_c^+}
\sqrt{2M_{\Sigma_c^{*\,++}} 2E_{\Lambda_c^+}(|\vec{q}\,|) }}
{|\vec{q}\,|\sqrt{2M_{\Sigma_c^{*\,++}} \left(
E_{\Lambda_c^+}(|\vec{q}\,|)+M_{\Lambda_c^+}\right)}}
\left(
(M_{\Sigma_c^{*\,++}}- E_{\Lambda_c^+}(|\vec{q}\,|) )\ A_{\Sigma_c^{*\,++}
\Lambda_c^+
,\ 0}^{1/2,1/2}
+|\vec{q}\,|\ A_{\Sigma_c^{*\,++}\Lambda_c^+
,\ 3}^{1/2,1/2}
\right)
\end{eqnarray}
with
\begin{eqnarray}
A_{\Sigma_c^{*\,++}\Lambda_c^+
,\ \mu}^{1/2,1/2}={}_{\stackrel{}{NR}}\left\langle \Lambda_c^+,\
1/2\ -|\vec{q}\,|\,\vec{k}\,\left|\  J_{A\,\mu}^{d\, u}(0)\ \right|
\,\Sigma_c^{*\,++},\ 1/2\ \vec{0}\,\right\rangle_{NR,\ non-pole}
\end{eqnarray}
The expressions for $A_{\Sigma_c^{*\,++}\Lambda_c^+
,\ \mu}^{1/2,1/2}$ ($\mu=0,3$) can be found in the appendix.\\
Results for $g_{\Sigma_c^{*\,++}\Lambda_c^+\pi^+}$ and the total widths 
$\Gamma(\Sigma_c^{*\,++}\to\Lambda_c^+\pi^+)$, 
$\Gamma(\Sigma_c^{*\,+}\to\Lambda_c^+\pi^0)$ and
$\Gamma(\Sigma_c^{*\,-}\to\Lambda_c^+\pi^-)$
appear in Table~\ref{tab:sig*lam}. Our value for the latter two are obtained with the use
of
$g_{\Sigma_c^{*\,++}\Lambda_c^+\pi^+}$ and with the appropriate mass
changes in the rest of  factors in Eq.~(\ref{eq:gammasig*lam}).
Our central value for $\Gamma(\Sigma_c^{*\,++}\to\Lambda_c^+\pi^+)$ is above 
the central value of the
latest experimental determination by the CLEO Collaboration in 
Ref.~\cite{cleo05}. For some of the potentials used, AP1 and AP2 of
Ref.~\cite{silvestre96}, the results  obtained are within experimental errors. The central value for
$\Gamma(\Sigma_c^{*\,+}\to\Lambda_c^+\pi^0)$ is
slightly above the upper experimental bound determined also by the 
CLEO Collaboration in Ref.~\cite{cleo01}, but again, we obtain results 
which are below the experimental bound using the AP1 and AP2 potentials.
  As for  $\Gamma(\Sigma_c^{*\,-}\to\Lambda_c^+\pi^-)$
we get a nice agreement with experiment. Our results for the different charge
configurations differ by 4\%
 at most. With the new value for $M_{\Lambda_c^+}$ given by the {\sl BABAR} 
Collaboration in Ref.~\cite{babar05} they would get reduced by 3\%. 
 Our results are
globally in better agreement with experiment than the ones obtained by other
 theoretical
calculations~\footnote{We did not show those cases were data on
$\Sigma_c^*\to\Lambda_c\pi$ widths were used to fit parameters of the models.}
with perhaps the exception of the QCDSR calculation of Ref.~\cite{zhu98}.
\subsection{$\Xi_c^{*}\to \Xi_c\pi$ decay}
\label{sect:xi*}
Once more we analyze  the case with a $\pi^+$ in the final state, 
$\Xi_c^{*\,+}\to \Xi_c^0\pi^+$.
 What we obtain is
\begin{eqnarray}
\label{eq:amcas}
{\cal A}^{(s,s')}_{\Xi_c^{*\,+}\Xi_c^0\pi^+}(P,P')&=&\frac{-i}{f_\pi}
\left\langle \Xi_c^0,\ s'\,\vec{P}'\,|\,q^\mu\, J_{A\,\mu}^{d\, u}(0)\,|
\,\Xi_c^{*\,+},\ s\,\vec{P}\,\right\rangle_{non-pole}\nonumber\\
&=&i\frac{g_{\Xi_c^{*\,+}\Xi_c^+\pi^+}}{M_{\Xi_c^+}+M_{\Xi_c^0}}
\ q_\nu\ \overline{u}_{\Xi_c^0\ s'}(\vec{P}'\,)\
u_{\Xi_c^{*\,+}\ s}^\nu(\vec{P})
\end{eqnarray}
where again we have introduced a dimensionless coupling 
 $g_{\Xi_c^{*\,+}\Xi_c^0\pi^+}$.
The width is given as
\begin{equation}
\label{eq:gammacas*cas}
\Gamma(\Xi_c^{*\,+}\to\Xi_c^0\pi^+)=\frac{|\vec{q}\,|^3}{24\pi M_{\Xi_c^{*\,+}}^2 }\
\frac{g_{\Xi_c^{*\,+}\Xi_c^0\pi^+}^2}{(M_{\Xi_c^{+}}+M_{\Xi_c^{0}})^2} \left((
M_{\Xi_c^{*\,+}}+M_{\Xi_c^0})^2-m^2_{\pi}\right)
\end{equation}
\begin{table}[t]
\begin{center}
\begin{tabular}{l|c c c c c}
\hline
 &\hspace{.01cm} $\ g_{\Xi_c^{*\,+}\Xi_c^0\pi^+}$\hspace{.01cm} 
 &\hspace{.01cm} $\Gamma(\Xi_c^{*\,+}\to\Xi_c^0\pi^+)$ \hspace{.01cm} 
 &\hspace{.01cm} $\Gamma(\Xi_c^{*\,+}\to\Xi_c^+\pi^0)$ \hspace{.01cm} 
 &\hspace{.01cm} $\Gamma(\Xi_c^{*\,0}\to\Xi_c^+\pi^-)$ \hspace{.01cm}
 &\hspace{.01cm} $\Gamma(\Xi_c^{*\,0}\to\Xi_c^0\pi^0)$ \hspace{.01cm} \\
& &[MeV] & [MeV]&[MeV]&[MeV]\\ \hline
 & &&&&\\
This work  
&$-28.83\pm0.50\pm0.10$ &$1.84\pm0.06\pm0.01$&
$1.34\pm0.04\pm0.01$&$2.07\pm0.07\pm0.01$&$0.956\pm0.030\pm0.007$\\
\hline
Theory  &&&&\\
LFQM  & &$1.12$~\cite{tawfiq98} & $0.69$~\cite{tawfiq98} &
  $1.16$~\cite{tawfiq98}& $0.72$~\cite{tawfiq98}\\
RTQM& &$1.78\pm0.33$~\cite{ivanov9899} &$1.26\pm0.17$~\cite{ivanov9899}&$2.11\pm0.29$~\cite{ivanov9899}
  & $1.01\pm0.15$~\cite{ivanov9899} \\
\hline
&&&&&\\
\multicolumn{1}{l|}{}&&\multicolumn{2}{c}
{$\Gamma(\Xi_c^{*\,+}\to\Xi_c^0\pi^++\Xi_c^+\pi^0)$}
&\multicolumn{2}{c}{$\Gamma(\Xi_c^{*\,+}\to\Xi_c^0\pi^++\Xi_c^+\pi^0)$}\\
 &&\multicolumn{2}{c}{[MeV]}
&\multicolumn{2}{c}{[MeV]}\\
\hline
\multicolumn{1}{l|}{This work}&&\multicolumn{2}{c}{$3.18\pm0.10\pm0.01$}&
\multicolumn{2}{c}{$3.03\pm0.10\pm0.01$}\\
\hline
\multicolumn{1}{l|}{Experiment}& & \multicolumn{2}{c}
{$<3.1$ (CL=90\%)~\cite{cleo96}}&\multicolumn{2}{c}{$<5.5$ 
(CL=90\%)~\cite{cleo95}}\\
\hline
Theory  &&&&\\
CQM& & \multicolumn{2}{c}{$<2.3\pm0.1$~\cite{rosner95}}  &
\multicolumn{2}{c}{$<2.3\pm0.1$~\cite{rosner95}}\\
 & & \multicolumn{2}{c}{1.191\,--\, 3.971~\cite{pirjol97}} & 
 \multicolumn{2}{c}{1.230\, --\, 4.074~\cite{pirjol97}}\\
 HHCPT && \multicolumn{2}{c}{$2.44\pm0.85$~\cite{cheng97}} &
  \multicolumn{2}{c}{$2.51\pm0.88$~\cite{cheng97}}\\
LFQM & & \multicolumn{2}{c}{1.81~\cite{tawfiq98}} &
 \multicolumn{2}{c}{1.88~\cite{tawfiq98}}\\
RTQM & &\multicolumn{2}{c}{$3.04\pm0.50$~\cite{ivanov9899}} &
 \multicolumn{2}{c}{$3.12\pm0.33 $~\cite{ivanov9899}}\\
 \hline
\end{tabular}
\end{center}
\caption{ Values for  the coupling $g_{\Xi_c^{*\,+}\Xi_c^0\pi^+}$ and 
decay widths $\Gamma(\Xi_c^{*\,+}\to\Xi_c^0\pi^+)$,
$\Gamma(\Xi_c^{*\,+}\to\Xi_c^+\pi^0)$,
$\Gamma(\Xi_c^{*\,0}\to\Xi_c^+\pi^-)$ and
$\Gamma(\Xi_c^{*\,0}\to\Xi_c^0\pi^0)$.
Experimental upper bounds for the total $\Xi_c^{*\,+}$ and 
$\Xi_c^{*\,0}$ widths, and different theoretical
calculations are also shown.}
\label{tab:cas*cas}
\end{table}
Taking now
$\vec{P}=\vec{0}$,  $\vec{P}'=-|\vec{q}\,|\,\vec{k}$ in the $z$ direction,
and $s=s'=1/2$, $g_{\Xi_c^{*\,+}\Xi_c^0\pi^+}$ is evaluated from Eq.(\ref{eq:amcas}) to be
\begin{eqnarray}
\label{eq:gcas*cas}
\hspace{-0.5cm}g_{\Xi_c^{*\,+}\Xi_c^0\pi^+}
&=&\frac{\sqrt3}{f_\pi\sqrt2}\ \frac{(M_{\Xi_c^+}+M_{\Xi_c^0})
\sqrt{2M_{\Xi_c^{*\,+}} 2E_{\Xi_c^0}(|\vec{q}\,|) }}
{|\vec{q}\,|\sqrt{2M_{\Xi_c^{*\,+}} \left(
E_{\Xi_c^0}(|\vec{q}\,|)+M_{\Xi_c^0}\right)}}
\left(
(M_{\Xi_c^{*\,+}}- E_{\Xi_c^0}(|\vec{q}\,|) )\ A_{\Xi_c^{*\,+}
\Xi_c^0
,\ 0}^{1/2,1/2}
+|\vec{q}\,|\ A_{\Xi_c^{*\,+}\Xi_c^0
,\ 3}^{1/2,1/2}
\right)
\end{eqnarray}
with
\begin{eqnarray}
A_{\Xi_c^{*\,+}\Xi_c^0
,\ \mu}^{1/2,1/2}={}_{\stackrel{}{NR}}\left\langle \Xi_c^0,\
1/2\ -|\vec{q}\,|\,\vec{k}\,|\  J_{A\,\mu}^{d\, u}(0)\ |
\,\Xi_c^{*\,+},\ 1/2\ \vec{0}\,\right\rangle_{NR,\ non-pole}
\end{eqnarray}
which expressions can be found in the  appendix.\\
 Results for the coupling $g_{\Xi_c^{*\,+}\Xi_c^0\pi^+}$, 
the widths $\Gamma(\Xi_c^{*\,+}\to\Xi_c^0\pi^+)$,
$\Gamma(\Xi_c^{*\,+}\to\Xi_c^+\pi^0)$,
$\Gamma(\Xi_c^{*\,0}\to\Xi_c^+\pi^-)$ and
$\Gamma(\Xi_c^{*\,0}\to\Xi_c^0\pi^0)$,  and the total
  widths  
$\Gamma(\Xi_c^{*\,+}\to\Xi_c^0\pi^++\Xi_c^+\pi^0)$  and
$\Gamma(\Xi_c^{*\,0}\to\Xi_c^0\pi^0+\Xi_c^+\pi^-)$ 
appear in Table~\ref{tab:cas*cas}. Our values for 
$\Gamma(\Xi_c^{*\,+}\to\Xi_c^+\pi^0)$,
$\Gamma(\Xi_c^{*\,0}\to\Xi_c^+\pi^-)$ and
$\Gamma(\Xi_c^{*\,0}\to\Xi_c^0\pi^0)$
are obtained with the use of $g_{\Xi_c^{*\,+}\Xi_c^0\pi^+}
/(M_{\Xi_c^+}+M_{\Xi_c^0})$, and with the appropriate mass changes
in the rest of  factors in Eq.~(\ref{eq:gammacas*cas}).
For $\Gamma(\Xi_c^{*\,+}\to\Xi_c^+\pi^0)$ and
$\Gamma(\Xi_c^{*\,0}\to\Xi_c^0\pi^0)$ an extra $1/2$ isospin factor 
should be included.
Our results for $\Gamma(\Xi_c^{*\,+}\to\Xi_c^0\pi^++\Xi_c^+\pi^0)$ are slightly
above the experimental bound obtained by the CLEO Collaboration~\cite{cleo96}.
As for the $\Sigma_c^*$ decay  case above, the AP1 and AP2 potentials gives results closer 
to experiment. For $\Gamma(\Xi_c^{*\,0}\to\Xi_c^+\pi^-+\Xi_c^0\pi^0)$ our result
is well below the CLEO Collaboration experimental bound in Ref.~\cite{cleo95}.
Isospin breaking due to mass effects is clearly seen when comparing the
predictions for $\Gamma(\Xi_c^{*\,+}\to\Xi_c^0\pi^-)$ and
$\Gamma(\Xi_c^{*\,+}\to\Xi_c^+\pi^0)$. One finds a factor 1.4 difference when
 a factor of two would be 
expected from isospin symmetry. Again, the fact that there is little phase space
available makes the results very sensitive to the actual mass values.
Compared to other theoretical calculations our results agree nicely with the
ones obtained within the RTQM of Ref.~\cite{ivanov9899}, while they are larger
than most other  determinations.

\section{Concluding remarks}
\label{sect:conclusions}
We have evaluated
the widths for the charmed-baryon decays
$\Sigma_c^*\to\Lambda_c\pi$,
$\Sigma_c\to\Lambda_c\pi$ and $\Xi_c^*\to\Xi_c\pi$ within the framework of a 
nonrelativistic quark model. While there have been dynamical calculations of
these reactions in other models, to our knowledge this is the first time  that
such a calculation has been  done within a nonrelativistic approach. We have used wave functions 
constrained by HQS and that
were obtained solving the nonrelativistic three--body problem with the help of a simple variational ansazt.
For that purpose we took five different nonrelativistic quark--quark 
interactions that included a
confining term plus Coulomb and hyperfine terms coming from one-gluon exchange.
To evaluate the pion
emission amplitude we have used a spectator or one-quark pion emission model. The
amplitude has been  obtained with the use of PCAC from the analysis of weak current
matrix elements.  Our results  are rather stable against 
the quark--quark interaction,  with variations in the decay widths 
 at the level of 6--8\%. We find an overall good agreement
with experiment for the three reactions. This agreement  is,
 in most cases, better than the one obtained by other models.
\begin{acknowledgments}
This research was supported by DGI and FEDER funds, under contracts
BFM2002-03218, BFM2003-00856 and  FPA2004-05616,  by the Junta de Andaluc\'\i a and
Junta de Castilla y Le\'on under contracts FQM0225 and
SA104/04, and it is part of the EU
integrated infrastructure initiative
Hadron Physics Project under contract number
RII3-CT-2004-506078.  C. A. wishes to acknowledge a grant  
from  Universidad de Granada and Junta de Andaluc\'\i a. 
J.\ M. V.-V. acknowledges a grant
(AP2003-4147) from the
Spanish Ministerio de Educaci\'on y Ciencia.
\end{acknowledgments}

\appendix
\section{Expressions for the $A^{1/2,1/2}_{BB',\, \mu}$  matrix elements}
\label{app}
The values for the $A^{1/2,1/2}_{BB',\, \mu}$  are evaluated using one-body current
operators and their expressions are given by
\begin{eqnarray}
A^{1/2,1/2}_{\Sigma_c^{++}\Lambda_c^+,\, \mu}
&=&\frac{\sqrt2}{\sqrt3}
\int d^{\,3}Q_1\ d^{\,3}Q_2\ 
\phi^{S_l=1}_{u,u,c}(\vec{Q}_1,\vec{Q}_2)
\left(
\phi^{S_l=0}_{d,u,c}(\vec{Q}_1-\frac{m_u+m_c}{\overline{M}_{\Lambda_c^+}}\, |
\vec{q}\,|\,\vec{k},\ \vec{Q}_2+\frac{m_u}{\overline{M}_{\Lambda_c^+}}\, |
\vec{q}\,|\,\vec{k} )
\right)^*\nonumber\\
&&\times\ \sum_{s_1}\,(1/2,1/2,1;s_1,-s_1,0)\,(1/2,1/2,0;s_1,-s_1,0)
\ \frac{
\overline{u}_{d\ s_1}(\vec{Q}_1-|
\vec{q}\,|\,\vec{k} )\ \gamma_\mu\gamma_5\ u_{u\, s_1}(\vec{Q}_1)}{\sqrt{2E_d(|\vec{Q}_1-|
\vec{q}\,|\,\vec{k} |)2E_u(|\vec{Q}_1|)}}\ 
\end{eqnarray}
%
%
%
where the quark Dirac spinors
are normalized to twice the energy. For $\mu=0,3$ we get the final 
expressions

\begin{eqnarray}
&&\hspace{-2cm}A^{1/2,1/2}_{\Sigma_c^{++}\Lambda_c^+,\, 0}
=\frac{\sqrt2}{\sqrt3}
\int d^{\,3}Q_1\ d^{\,3}Q_2\ 
\phi^{S_l=1}_{u,u,c}(\vec{Q}_1,\vec{Q}_2)
\left(
\phi^{S_l=0}_{d,u,c}(\vec{Q}_1-\frac{m_u+m_c}{\overline{M}_{\Lambda_c^+}}\, |
\vec{q}\,|\,\vec{k},\ \vec{Q}_2+\frac{m_u}{\overline{M}_{\Lambda_c^+}}\, |
\vec{q}\,|\,\vec{k} )
\right)^*\nonumber\\
&&\times\ \sqrt{\frac{\left(E_d(|\vec{Q}_1-|
\vec{q}\,|\,\vec{k} |)+m_d\right)\left(E_u(|\vec{Q}_1|)+m_u\right)}{{2E_d(|\vec{Q}_1-|
\vec{q}\,|\,\vec{k} |)2E_u(|\vec{Q}_1|)}}}\ 
\left(
\frac{Q_1^z}{E_u(|\vec{Q}_1|)+m_u}+
\frac{Q_1^z-|\vec{q}\,|}{E_d(|\vec{Q}_1-|
\vec{q}\,|\,\vec{k} |)+m_d}
\right)
\end{eqnarray}
\begin{eqnarray}
&&\hspace{-1.5cm}A^{1/2,1/2}_{\Sigma_c^{++}\Lambda_c^+,\, 3}
=-\frac{\sqrt2}{\sqrt3}
\int d^{\,3}Q_1\ d^{\,3}Q_2\ 
\phi^{S_l=1}_{u,u,c}(\vec{Q}_1,\vec{Q}_2)
\left(
\phi^{S_l=0}_{d,u,c}(\vec{Q}_1-\frac{m_u+m_c}{\overline{M}_{\Lambda_c^+}}\, |
\vec{q}\,|\,\vec{k},\ \vec{Q}_2+\frac{m_u}{\overline{M}_{\Lambda_c^+}}\, |
\vec{q}\,|\,\vec{k} )
\right)^*\nonumber\\
&&\hspace{.3cm}\times\ \sqrt{\frac{\left(E_d(|\vec{Q}_1-|
\vec{q}\,|\,\vec{k} |)+m_d\right)\left(E_u(|\vec{Q}_1|)+m_u\right)}{{2E_d(|\vec{Q}_1-|
\vec{q}\,|\,\vec{k} |)2E_u(|\vec{Q}_1|)}}}
\left(1+\frac{ 2(Q_1^z)^2-|\vec{Q}_1|^2-Q_1^z|
\vec{q}\,|}{\left(E_d(|\vec{Q}_1-|
\vec{q}\,|\,\vec{k} |)+m_d\right)\left(E_u(|\vec{Q}_1|)+m_u\right)}
\right)
\end{eqnarray}
\vspace{1cm}\\
For $A^{1/2,1/2}_{\Sigma_c^{*\,++}\Lambda_c^+,\, \mu}$ we just have
\begin{equation}
A^{1/2,1/2}_{\Sigma_c^{*\,++}\Lambda_c^+,\, \mu}
=-\sqrt2\ A^{1/2,1/2}_{\Sigma_c^{++}\Lambda_c^+,\, \mu}
\end{equation}
%
%
%
Similarly we get for $A^{1/2,1/2}_{\Xi_c^{*\, +}\Xi_c^0,\, \mu},\ \mu=0,3$
\begin{eqnarray}
&&\hspace{-2cm}A^{1/2,1/2}_{\Xi_c^{*\, +}\Xi_c^0,\, 0}
=\frac{\sqrt2}{\sqrt3}
\int d^{\,3}Q_1\ d^{\,3}Q_2\ 
\phi^{S_l=1}_{u,s,c}(\vec{Q}_1,\vec{Q}_2)
\left(
\phi^{S_l=0}_{d,s,c}(\vec{Q}_1-\frac{m_s+m_c}{\overline{M}_{\Xi_c^0}}\, |
\vec{q}\,|\,\vec{k},\ \vec{Q}_2+\frac{m_s}{\overline{M}_{\Xi_c^0}}\, |
\vec{q}\,|\,\vec{k} )
\right)^*\nonumber\\
&&\times\ \sqrt{\frac{\left(E_d(|\vec{Q}_1-|
\vec{q}\,|\,\vec{k} |)+m_d\right)\left(E_u(|\vec{Q}_1|)+m_u\right)}{{2E_d(|\vec{Q}_1-|
\vec{q}\,|\,\vec{k} |)2E_u(|\vec{Q}_1|)}}}\ 
\left(
\frac{Q_1^z}{E_u(|\vec{Q}_1|)+m_u}+
\frac{Q_1^z-|\vec{q}\,|}{E_d(|\vec{Q}_1-|
\vec{q}\,|\,\vec{k} |)+m_d}
\right)
\end{eqnarray}
\begin{eqnarray}
&&\hspace{-1.5cm}A^{1/2,1/2}_{\Xi_c^{*\,+}\Xi_c^0,\, 3}
=-\frac{\sqrt2}{\sqrt3}
\int d^{\,3}Q_1\ d^{\,3}Q_2\ 
\phi^{S_l=1}_{u,s,c}(\vec{Q}_1,\vec{Q}_2)
\left(
\phi^{S_l=0}_{d,s,c}(\vec{Q}_1-\frac{m_s+m_c}{\overline{M}_{\Xi_c^0}}\, |
\vec{q}\,|\,\vec{k},\ \vec{Q}_2+\frac{m_s}{\overline{M}_{\Xi_c^0}}\, |
\vec{q}\,|\,\vec{k} )
\right)^*\nonumber\\
&&\times\ \sqrt{\frac{\left(E_d(|\vec{Q}_1-|
\vec{q}\,|\,\vec{k} |)+m_d\right)\left(E_u(|\vec{Q}_1|)+m_u\right)}{{2E_d(|\vec{Q}_1-|
\vec{q}\,|\,\vec{k} |)2E_u(|\vec{Q}_1|)}}}
\ \left(1+\frac{2(Q_1^z)^2-|\vec{Q}_1|^2-Q_1^z|
\vec{q}\,|}{\left(E_d(|\vec{Q}_1-|
\vec{q}\,|\,\vec{k} |)+m_d\right)\left(E_u(|\vec{Q}_1|)+m_u\right)}
\right)
\end{eqnarray}
%
%
%
%
%

\end{document}